# Can the THz-TDS detect trace gases?

Noureddin OSSEIRAN
IEMN UMR 8520
CNRS, Univ. Lille
Villeneuve d'Ascq, France
noureddin.osseiran@univ-littoral.fr

Jeyan BICHON
IEMN UMR 8520
CNRS, Univ. Lille
Villeneuve d'Ascq, France
jeyan.bichon.etu@univ-lille.fr

Sophie ELIET
IEMN UMR 8520
CNRS, Univ. Lille
Villeneuve d'Ascq, France
sophie.eliet@univ-lille.fr

Romain PERETTI
IEMN UMR 8520
CNRS, Univ. Lille
Villeneuve d'Ascq, France
romain.peretti@univ-lille.fr

*Abstract*—The THz-TDS is a versatile technique and can be used to probe gas phase molecules. We are pushing the technique to its limit to try and detect trace gases at the sub-ppm levels. The experiments served as a reference for a quantitative approach that permits to extract the concentration of a probed gas. The single-parameter quantification model can be considered as the basis for a wider multispecies detection scheme.

## I. Introduction

THz Time Domain Spectroscopy (TDS) is a powerful technique that is widely recognized for characterizing a diverse range of materials, including semiconductors, liquids, biomolecules and meta-surfaces. It is less known for detecting gases despite its promising performance. The broadband capabilities of commercial terahertz time-domain spectroscopy (THz-TDS) systems (spanning 0.2 to 6 THz) make them particularly appealing for atmospheric and environmental applications that require multispecies detection and quantification in the gas phase. Another powerful aspect of THz-TDS is its high selectivity within the multispecies detection scheme [1]. On the other hand, the adoption of THz-TDS for gas-phase studies remains limited due to its shy performance when it comes to spectral resolution. This limitation primarily stems from the mechanical limitation of the delay line, also known as Fourier Heisenberg criteria (FHC), resulting in a typical resolution of approximately 1.2 GHz for THz-TDS systems. However, it has been shown recently that advanced numerical data treatment can achieve ~10-fold enhancement in resolution[2], which allows the TDS to match the high-resolution demands of gas phase studies.

An additional point to consider when thinking about the gas phase application of the THz-TDS is its sensitivity and in particular its limit of detection. Can it detect trace gases on the sub-ppm levels? These kinds of studies are scarce and this work is intended to open the doors toward trace gas detection and quantification. We chose to target Ammonia $NH_3$. Our choice is influenced by the fact that $NH_3$ has been widely studied, and also it was used to characterize our super-resolution algorithm before[3]. The interest in pushing the TDS to its limits arises from the fact that having high sensitivity can open the door towards a new field of applications. This is because having a highly sensitive and selective apparatus is critical for gas-phase multispecies detection protocols.

In this conference, we will propose an approach for the systematic retrieval of $NH_3$ concentrations from THz-TDS spectra based on experimental measurements at different concentrations.

## II. Experiments

The gas phase time traces were recorded by a MenloSystems TeraSmart TDS system. The THz radiation traversed a 1m vacuum gas cell using two 50 mm parabolic mirrors. Two TPX windows (2mm thickness) were used as the limits of the cell. Three Ammonia ($NH_3$) gas cylinders (AirLiquide) were used to prepare the different mixtures used throughout the experiments. Three sets of experiments were concluded as follows:

1. 1000 ppm $NH_3$ was diluted using N2 for concentrations of 1000-10 ppm
2. 10 ppm $NH_3$ was diluted using N2 for concentrations of 10- 0.05 ppm
3. 100% (pure) $NH_3$ was used as bought

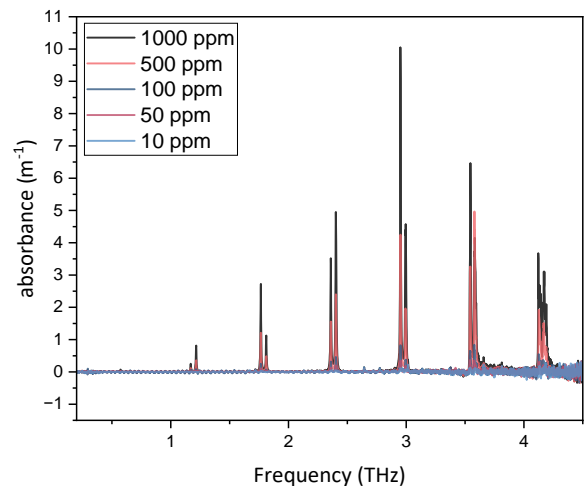

All the experiments were conducted at a total pressure of 1000 mbar.

Figure 1 Absorbance of $NH_3$ at different concentrations obtained by diluting 1000 ppm cylinder using $N_2$.

The dilutions were carried directly inside the vacuum gas chamber and using partial pressures.

## III. RESULTS

Figure 1 above shows the absorbance of NH$_3$ at different concentrations. The total pressure for all experiments is 1000 mbar.

The raw time traces were corrected and averaged using our home-written Python code Correct@TDS. The final average of 1000 time traces, for each experiment, was fitted in the time domain by the publicly available Fit@TDS software. The fit is a minimisation of an objective function containing all the parameters needed to describe the observed transitions with Lorentz oscillators. The fitted Lorentz parameters for all transitions will be used as input for a quantification model that will be explained after.

The absorbance shown in Figure 1 above is based on the transmission obtained by comparing the THz signal with and without a sample. Figure 2 below shows the preliminary results obtained for concentrations between 1 and 10 ppm. We extracted the area of two transitions of NH3 and plotted their evolution with concentration to demonstrate that we could go further down in concentration. More detailed results will be presented at the conference. Data analysis is currently being carried out to extract intensities and line widths below 1 ppm. Post-signal processing with error estimation is indispensable

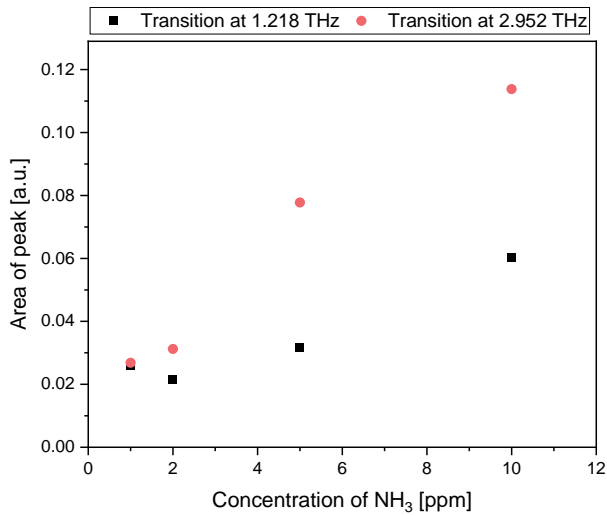

*Figure 2 Area of two peaks of NH$_3$ at 1.218 and 2.952 THz as function of concentration in ppm*

for trace analysis.

## IV. METHODOLOGY

Our vision of the current work falls within the context of maximizing the information extracted from THz-TDS data. Our group has introduced several approaches and algorithms that employ digital data processing to extend and minimize the uncertainties of experimental data[4], [5], [6]. Within this context, a new NH$_3$ gas model, based on the above experimental results, will be implemented in the Fit@TDS code. The implementation is in the form of a single-parameter model specific to NH$_3$ with the possibility of adding more gases in the future.

Starting from a Lorentz model for oscillators, we can describe the permittivity of the overall system by the following equation:

$$\tilde{\epsilon} = \tilde{\eta}^2(\omega) = \tilde{\epsilon}_\infty + \sum_{k=1}^{k_{max}} \frac{\Delta\epsilon_k \omega_{0k}^2}{\omega_{0k}^2 - \omega^2 + i\omega\gamma_k}$$

Where $\tilde{\epsilon}_\infty$ is the dielectric permittivity at high frequency compared to the range of interest, $k_{max}$ is the number of considered oscillators, $\Delta\epsilon_k$, $\omega_{0k}$ and $\gamma_k$ are the strength (units of permittivity), resonant frequency and damping rate of the kth oscillator, respectively. Keeping this equation in mind, we can say that the total signal depends on the contribution of all lines. The spectral line strength $\Delta\epsilon_k$ is proportional to the number of molecules probed, and in our case the partial pressure of NH$_3$. Based on this proportionality we can extract any partial pressure starting from a known reference, which we will take as our experimental pure Ammonia data at 1000 mbar. A new parameter that we call the pressure line strength $\Delta\zeta'_k$ can be defined as:

$$\Delta\zeta'_k = P_i \frac{\Delta\epsilon_{k,\text{ref}}}{P_{t,ref}}$$

Where $P_i$ is the partial pressure, $P_{t,ref}$ is the total pressure of reference (1000 mbar) and $\Delta\epsilon_{k,ref}$ is the oscillator strength at the ref concentration and pressure. Considering partial pressure is constant for a given experiment. The permittivity will become:

$$\tilde{\epsilon} = \tilde{\eta}^2(\omega) = \tilde{\epsilon}_\infty + \frac{P_i}{P_t}\sum_{k=1}^{k_{max}} \frac{\Delta\epsilon_{k,ref}\ \omega_{0k}^2}{\omega_{0k}^2 - \omega^2 + i\omega\gamma_k}$$

From here we can see that it is possible to fit any data to a reference spectrum of Ammonia to extract the concentration. The reference here is our proper 1000 mbar measurement.

The broadening or damping rate $\gamma_k$ depends on the total pressure, this is why the total pressure for all experiments should be similar for the model to give meaningful information.

It is also important to mention that the proposed model can be used with other models such as the water removing approach discussed elsewhere. Multi-model implementation can be advantageous in terms of eliminating any correlation present due to other molecules.

## V. CONCLUSION

We have demonstrated here the capabilities of THz-TDS to detect gas phase molecules in the ppm ranges. The current work opens the door to a myriad of applications that would make use of the simplicity of single-parameter models and their capabilities in multispecies detection and quantification.

In this conference, we will present our latest data treatment results for sub-ppm measurements as well.

## ACKNOWLEDGEMENT

The current work was funded by La Région Hauts-de-France and the French CNRS.